\documentclass[11pt,a4paper]{article}
\usepackage[]{amsmath,amssymb}
\usepackage{graphics,epsfig}
\begin{document}
\date{}
\title{Remarks on the entanglement entropy for disconnected regions}
\author{H. Casini\footnote{e-mail: casini@cab.cnea.gov.ar}
 \, and M. Huerta\footnote{e-mail: marina.huerta@cab.cnea.gov.ar} \\
{\sl Centro At\'omico Bariloche,
8400-S.C. de Bariloche, R\'{\i}o Negro, Argentina}}
\maketitle

\begin{abstract}
Few facts are known about the entanglement entropy for disconnected regions in quantum field theory. We study here the property of extensivity of the mutual information, which holds for free massless fermions in two dimensions. We uncover the structure of the entropy function in the extensive case, and find an interesting connection with the renormalization group irreversibility. The solution is a function on space-time regions which complies with all the known requirements a relativistic entropy function has to satisfy. We show that the holographic ansatz of Ryu and Takayanagi, the free scalar and Dirac fields in dimensions greater than two, and the massive free fields in two dimensions all fail to be exactly extensive, disproving recent conjectures. 
\end{abstract}

\section{Introduction}
The entanglement entropy enclosed in a region $V$ of the space is ambiguous in quantum field theory (QFT). This is because there is a divergent amount of entanglement between the $V$ and the exterior, due to the vacuum fluctuations. 
However, as it happens with the Casimir effect, the unwanted effects of these fluctuations can be governed by looking at different, cutoff independent quantities, which can be constructed from the local entropy.
 With this aim let us introduce the mutual information between two non intersecting regions $A$ and $B$, defined by 
\begin{equation}
I(A,B)=S(A)+S(B)-S(A\cup B)\,,\label{dos}
\end{equation}  
where the entanglement entropy $S(V)=-\textrm{tr} \rho_V \log \rho_V$ is the the von Neumann entropy of the local density matrix $\rho_V$ corresponding to the global vacuum state. 
Since the divergences in $S(V)$ are related to the boundary of $V$, 
$I(A,B)$ is free from divergences and ambiguities, and thus a good measure of entanglement in QFT. 

The calculation of $I(A,B)$ requires the knowledge of the entanglement entropy for multicomponent sets, from which very little is known so far. In particular, the only exact result is for free massless fermions in two dimensions. For a Dirac field the multicomponent entropy function is  \cite{cteo,ch3},
\begin{equation}
S(X_1 \cup X_2 \cup ... \cup X_p)=\frac{1}{3}\left( \sum_{i,j} \log | a_i-b_j |-\sum_{i<j}
\log |a_i -a_j |-\sum_{i<j}\log | b_i -b_j |-p\,\log\epsilon \right)\,,\label{confor}
\end{equation}
where $a_i$ and $b_i$ are the left and right endpoints of the interval $X_i$ and $\epsilon$ is a short distance ultraviolet cutoff.
Remarkably, this specific form of the entropy gives an extensive mutual information \cite{ch3}, 
\begin{equation}
I(A,B\cup C)=I(A,B)+I(A,C)\,.\label{ex}
\end{equation}

Given this peculiar result, one may naturally wonder if there are other QFT with the same property. For a couple of years it was thought that the expression (\ref{confor}) for the entropy of the free massless Dirac field, multiplied by a global factor of the Virasoro central charge, was valid for any two dimensional conformal field theory (CFT) \cite{cc}. Recently, the derivation of this result was called into question (see \cite{cha1} and the note added to \cite{cc}), and up to the present, the expression of the entropy for more than one interval and more general CFT remains unknown. The work \cite{cha2} reports numerical results for the compactified scalar which hint to a different behavior than (\ref{confor}).

In this work, we will investigate the property of extensivity of the mutual information in QFT from a general point of view. The results allow us to disprove extensivity for some theories in more than two dimensions. In the course of the investigation we also find an unexpected connection with renormalization group irreversibility.

Two recent developments add interest to this subject. The first one is the geometric ansatz for the entanglement entropy of a CFT proposed by Ryu and Takayanagi in the context of the Maldacena duality \cite{rt}. This gives the entanglement entropy of a connected spatial region $V$ on the Minkowskian boundary of an AdS space as the area of a minimal surface in the bulk, whose boundary coincides with the one of $V$. This holographic entropy has been supported by several arguments \cite{sup,hht}. However, this construction does not provide a formula for the entropy corresponding to multicomponent sets. In \cite{hr} Hubeny and Rangamani propose one such extension (see also \cite{rt}). For the two dimensional case it is a generalization of (\ref{confor}) 
\begin{equation}
S(X_1 \cup X_2 \cup ... \cup X_p)=\sum_{i,j} S_1( | a_i-b_j |)-\sum_{i<j}
S_1( | a_i -a_j |)-\sum_{i<j}S_1(| b_i -b_j |)\,,\label{confora}
\end{equation}
where $S_1(l)$ is the entropy function for single component sets. This formula leads also to extensivity of the mutual information. For more dimensions the proposal in \cite{hr} gives an extensive $I(A,B)$ at least for a particular relative position of the sets involved. 
  
A different kind of investigation has also met with the present problem \cite{bh}. There, we have found that an unambiguous definition of the entropy contained in the Hawking radiation on a finite region $V$ is given by half the mutual information between the black hole and $V$. We have also argued using the mutual information that, in case there exists a Hagedorn transition, there is a natural regularization of the entanglement entropy in QFT which does not requires quantum gravity. This may hint to a different understanding of the finitness of the black hole entropy in terms of the maximum ammount of information that can be shared between the inside and the outside of the horizon. The interpretation of the ordinary thermodynamical radiation entropy (the entropy in the Hawking radiation) as mutual information requires the extensivity of this later at least in the particular limit case when $V$ is large and far from the black hole. The conditions for this to hold are unknown at present, and it also motivates further investigations on the properties of the mutual information in QFT.     

\section{General considerations}

The fact that $I(A,B)$ is a well defined quantity, and that we take the vacuum as global state, has the added advantage that the spatial symmetries of the theory (Poincar\'e, conformal) are explicitly manifest. 

Further relations come from causality \cite{hh}. 
Indeed, heuristically one would expect that the evolution between the Cauchy slices $A$ and $A^\prime$ in figure 1 is given by a unitary operator in quantum field theory. This would imply that the entropy is the same for these surfaces, $S(A)=S(A^\prime)$. In other terms, the causal evolution does not allow the information present in $A^\prime$ to move at a velocity greater than the light speed and then not to be present also in $A$ (escaping from the diamond shaped set $\hat{A}$ in figure 1). The information content must then remain constant between these sets. There is a subtle point however, since the entanglement entropy requires regularization and this may spoil the unitarity of the evolution. However, this problem is only apparent since it is solved (at the expense of dealing with two sets rather than one) by the use of the mutual information, which is regularization independent. The mutual information can be defined (at the full rigor of the axiomatic level in QFT) using only the global state and the algebras of the operators pertaining to two different regions $A$ and $B$. These later are exactly the same as the algebras of local operators corresponding to $A^\prime$ and $B^\prime$ due to the unitary causal evolution of the Heisemberg operators. Thus, these algebras, and the mutual information, are in fact funcions of the "diamond shaped" or causally complete sets  \cite{haag}.  These $D$ dimensional sets, where $D$ is the space-time dimension, are the domain of dependence of spatial sets and in this sense they are classes of equivalence of Cauchy slices which have the same domain of dependence. 
 The causality is then explicitly manifest in the mutual information leading to  
 $I(A,B)=I(A^\prime,B^\prime)$  whenever $A$ and $A^\prime$, and $B$ and $B^\prime$ are Cauchy surfaces for the same $D$ dimensional sets $\hat{A}$ and $\hat{B}$.  
We note that $A$ and $B$ have to be at a space-like distance in order to define a mutual information, since the respective operator algebras have to be commuting.

One can naturally wonder if all the information about universal (regularization independent) quantities contained in the entropy are already present in the function $I(A,B)$. Though it is not a theorem, a positive answer may be expected. This is because, due to the purity of the vacuum state, we have $S(A)=S(-A)$, where $-A$ is the set complementary to $A$, and the entropy corresponding to the whole space $A\cup -A$ is zero (again due to the purity of the vacuum $\left|0\right>$). Thus, we recover the entropy function from the mutual information by taking the limit 
 \begin{equation}
\lim_{B\rightarrow -A} I(A,B)=2 S(A)\,.\label{in}
\end{equation}  
In this limit the separation between $A$ and $B$  provides a cutoff, and the mutual information diverges when $A$ and $B$ come into contact, corresponding to a divergent $S(A)$.

Some properties of $I(A,B)$ can be deduced from its relation (\ref{dos}) with the entropy. The strong subadditivity of the entropy \cite{ssa}
\begin{equation}
S(A)+S(B)\ge S(A\cup B) +S(A\cap B)\,,\label{ssaa}
\end{equation}
implies that the mutual information is positive and monotonically increasing 
\begin{eqnarray}
I(A,B)&\ge& 0 \,,\\
 I(A,B)&\le& I(A,C), \hspace{2.5cm} B\subseteq C \,.\label{ine1}
\end{eqnarray} 

\begin{figure} [tbp]
\centering
\leavevmode
\epsfysize=3.7cm
\bigskip
\epsfbox{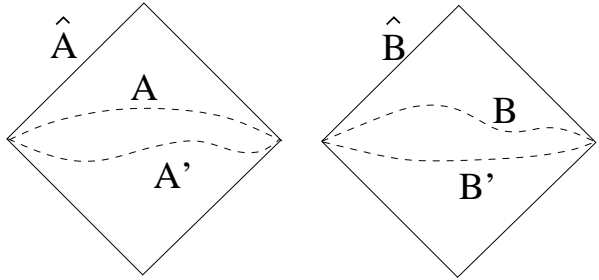}
\caption{Two spatial sets $A$ and $B$ in two dimensional Minkowski space-time. Light rays are plotted at $\pm 45^\circ$. The spatial sets $A$, $A^\prime$, and $B$, $B^\prime$, have the same causal domain of dependence $\hat{A}$ and $\hat{B}$ respectively. Then we have $I(A,B)=I(A^\prime,B^\prime)$. }
\end{figure}
 
 In order to measure the degree of extensivity of the mutual information we can introduce the tripartite information function (somehow information shared between $B$ and $C$ with respect to $A$)
\begin{equation}
I(A,B,C)=I(A,B)+I(A,C)-I(A,B\cup C)\,.\label{non}
\end{equation}
This function has complete permutation symmetry since 
\begin{equation}
I(A,B,C)\equiv S(A)+S(B)+S(C)-S(A\cup B)-S(A\cup C)-S(B\cup C)+S(A\cup B \cup C)\,.
\end{equation}
For extensive functions (i.e. (\ref{confor})) we have 
\begin{equation}
I(A,B,C)=0\,.
\end{equation}
One particular case when $I(A,B,C)=0$ is when the joint state for the three systems, $\rho_{A\cup B\cup C}$ is pure, a situation which does not hold in QFT unless that $A\cup B \cup C$ is the whole space. The general condition which makes the mutual information extensive for a tripartite system, $I(A,B,C)=0$, are unknown at present.

A particular limit of the tripartite information function is known to provide an order parameter for massive theories with topological order \cite{pres}. In this case $I(A,B,C)$ has been called topological entanglement entropy. In three dimensions this limit of $I(A,B,C)$ is always negative or zero, a fact that follows from the strong subadditive property.  

One may be tempted to speculate that a natural generalization of the strong subadditive inequality (\ref{ssaa}) for tripartite systems is that $I(A,B,C)$ has a definite sign. Unfortunately, this does not hold in general, and both signs are allowed  \cite{sign}. Below we show some examples of this fact in QFT.  

\section{Extensivity and the renormalization group irreversibility}

We now investigate the properties of theories which are extensive, that is, theories such that the mutual information satisfies the property (\ref{ex}), or equivalently $I(A,B,C)=0$. If the mutual information is extensive, $I(A,B)$ can be decomposed in a sum of the mutual information of infinitesimal sets covering the surfaces $A$ and $B$. Moreover, the result must be the same independently of the way in which this decomposition into small sets is made. Thus, modulo details from meassure theory, we can write the mutual information in terms of an integral
\begin{equation}
I(A,B)=\int_{A} d\sigma_x \,\, \int_{B} d\sigma_y\,\, j(x,y,\eta_{A,x}, \eta_{B,y})\,,\label{dfdf}
\end{equation}
where $\eta_A$ and $\eta_B$ are the normal vectors to the surfaces $A$ and $B$. Note that $j(x,y,\eta_{A,x},\eta_{B,y})$ in this expression depends on the points $x$, $y$, which can be thought as the location of the infinitesimal spatial sets, and their orientation in spacetime, given by the normal vectors $\eta_{A,x}$ and $\eta_{B,y}$ at these points.  However, $j(x,y,\eta_{A,x},\eta_{B,y})$ in equation (12) does not depend on other details of $A$ and $B$ since the same infinitesimal sets can be used to cover (and compute the mutual information of) any other pair of sets which passes through the same points $x$ and $y$ with the same normal orientations $\eta_{A,x}$ and $\eta_{B,y}$.

Causality requires that $I(A,B)$ must be independent of the Cauchy surfaces $A$ and $B$ for the causal completion $\hat{A}$ and $\hat{B}$   (the diamond shaped sets in figure 1). Now, one can make an infinitesimal transformation of the Cauchy surfaces in such a way that the points $x$ and $y$ are varied infinitesimally, but the normal vectors $\eta_{A,x}$ and  $\eta_{B,y}$ get macroscopically varied. In order that this type of ripples in the surfaces do not change the result of the integral in (\ref{dfdf}) the function $j(x,y,\eta_{A,x}, \eta_{B,y})$ must be linear in $\eta_A$ and $\eta_B$. Equation (\ref{dfdf}) then becomes  
\begin{equation}
I(A,B)=\int_{A} d\sigma_x \,\,\eta^\mu_{A,x} \, \int_{B} d\sigma_y\,\, \eta^\nu_{B,y}\, J_{\mu\nu}(x,y)\,,
\label{pasa}
\end{equation}
for certain symmetric tensor $J_{\mu\nu}(x,y)$ which is a function of the points $x$ and $y$ only.

Now, insisting in that the result of (\ref{pasa}) must not depend on arbitrary variations of the surfaces $A$ and $B$ (while keeping them spatial and their boundary fixed), we can use the Gauss theorem in order to prove that this bi-current must be conserved, 
\begin{equation}
\partial_\mu J^{\mu\nu}(x,y)=0\,.
\end{equation}

The Poincar\'e invariance then gives
\begin{equation}
J^{\mu\nu}(x,y)=\frac{(x-y)^\mu (x-y)^\nu}{(x-y)^{2 D}} G(\sqrt{(x-y)^2})-\frac{g^{\mu\nu}}{(x-y)^{2 (D-1)}} \, F(\sqrt{(x-y)^2})\,.\label{jota}
\end{equation}
We take the metric signature such that $(x-y)^2> 0$ for a pair of space-like separated points. The functions $F(l)$ and $G(l)$ are dimensionless. 
The conservation of $J^{\mu\nu}(x-y)$ leads to
\begin{equation}
(G-F)^\prime=-(D-1) \frac{(2 F-G)}{l}\,.\label{to}
\end{equation}
In particular, for an extensive conformal theory $F$ and $G$ must be constant, what according to (\ref{to}) means $G=2 F$, in any dimension. 

Another constraint to $F$ and $G$ follows from the positivity of the mutual information, which in the present extensive case also entails monotonicity. This implies
\begin{equation}
\eta^1_\mu J^{\mu\nu}(x,y) \eta^2_\nu\ge 0\,,
\end{equation}
for any vectors $\eta^1$ and $\eta^2$ in the positive light cone. This is equivalent to 
\begin{equation}
 2 F\ge  G \ge 0\,.
\end{equation}
Then, according to (\ref{to}), the dimensionless function $C(l)$, 
\begin{equation} 
C(l)=G(l)-F(l)
\end{equation}
 is decreasing 
\begin{equation}
C'(l)\le 0\,.
\end{equation}
This means there is a monotonous behavior along the renormalization group trajectories for extensive theories in any dimensions, a result resembling the $C$ theorem in two dimensions \cite{zamo}. At the fixed points it is $C=F\ge 0$. As we discuss in Section 4, in four dimensions, this quantity at the conformal point is proportional to the type $a$ conformal anomaly.  Then the $C^\prime (l)<0$ is consistent with the proposal that this anomaly plays the role of the Zamolodchikov's c-function in four dimensions \cite{cardy}.

The existence of an infrared fixed point then means that $G$ cannot get to be smaller than $F$, 
\begin{equation}
C(l)\ge 0\,
\end{equation}
for any $l$, and $\lim_{l\rightarrow \infty} (F(l)-C(l))=0$. From (\ref{to}) we have
\begin{eqnarray}
F=-(D-1)^{-1} \,l\,C^\prime + C\,,\\
G=-(D-1)^{-1} \,l\,C^\prime +2 C\,.
\end{eqnarray}
Thus, taking into account these equations, all the constraints are summarized by $C(l)$ being positive and decreasing. 
 
Writing 
\begin{equation}
C(l)=(D-1) l^{2D-3} H^\prime(l)\,
\end{equation}
for some function $H(l)$, we have 
\begin{equation}
J_{\mu\nu}(l)=-\partial_\mu \partial_\nu \,H(l) + g_{\mu\nu} \, \partial_\alpha \partial^\alpha  H(l)\,.
\end{equation}
The conservation of the current is transparent in this expression. With this we can integrate eq. (\ref{pasa}) to obtain
\begin{equation}
I(A,B)=-\int_{\partial A} d\sigma_x^{\alpha\beta}\, \int_{\partial B} d\sigma^y_{\alpha\beta}\, H(|x-y|)   \,,\label{h0}
\end{equation} 
where the integration is now over the $(D-2)$ dimensional boundaries of the spatial regions $A$ and $B$, taken with the same orientation. 
 In particular, when $A$ and $B$ are contained in the same spatial hyperplane it is
 \begin{equation}
 I(A,B)=\int_{\partial A} d\sigma_x \, \int_{\partial B} d\sigma_y \, H(|x-y|)  \,\, (\vec{\eta}^A_x . \vec{\eta}^B_y) \,,\label{h}
 \end{equation}
with $\vec{\eta}^A$ and $\vec{\eta}^B$ the outward pointing unit vectors normal to the surfaces $\partial A$ and $\partial B$ respectively. 
These boundaries depend on the causally complete regions $\hat{A}$ and $\hat{B}$, and are explicitly independent of the particular Cauchy surfaces. The arbitrary additive integration constant in $H$ is irrelevant since the surfaces $\partial A$ and $ \partial B$ are closed.

In the case of a conformal theory, the function $H(l)$ has the particular form 
\begin{equation}
H(l)=-\frac{C_D}{l^{2D-4}}\,,
\end{equation}  
where $C_D=C(0)/(2(D-1)(D-2))$, with $C(0)$ the constant value of the function $C(l)$ at the conformal point. In this case it can be checked that (\ref{pasa}) is invariant under the full conformal group in $D$ dimensions.

\section{Testing extensivity in different theories}

\begin{figure} [tbp]
\centering
\leavevmode
\epsfysize=4.3cm
\bigskip
\epsfbox{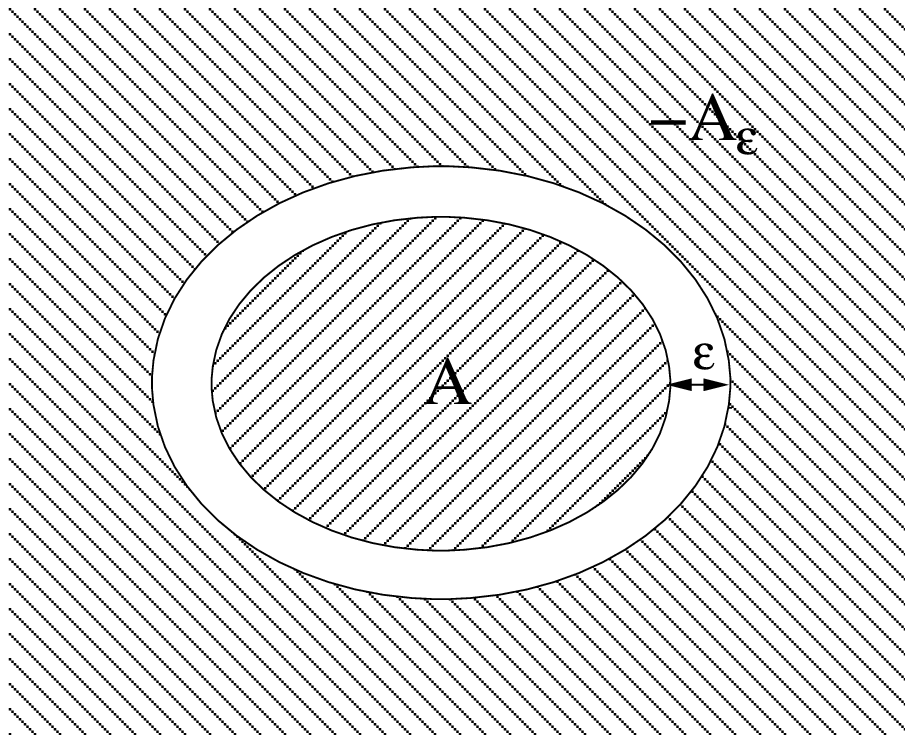}
\caption{In order to obtain the entropy corresponding to $A$, we can evaluate the mutual information between $A$ and $-A_\epsilon$, which is the set formed by the points separated from $A$ more than a short distance cutoff $\epsilon$.}
\end{figure}

In this Section we compare the general formula (\ref{h}) for the extensive case with some known exact results for specific theories. These are available in few cases, and correspond to free scalars and fermions \cite{a,b,c}, or to the CFT in the Ryu and Takayanagi holographic ansatz.

In order to make this comparison we take advantage of the property (\ref{in}), which expresses the entanglement entropy as a particular limit of the mutual information, and holds for pure states. In this Section we consider sets included in a single spatial hyperplane, and thus we can just impose a distance cutoff $\epsilon$
\begin{equation}
 S(A)=\frac{1}{2} \,I(A,-A_\epsilon)\,,\label{reg}
\end{equation}
where $-A_\epsilon$ is the set of points on the hyperplane of $A$ which are at distance longer than $\epsilon$ from $A$ (see figure 2). Expanding for $\epsilon\rightarrow0$ we get the entropy with its divergent and finite terms.

\subsection{Two dimensional case}   

For the extensive case in two dimensions we derive the entropy from (\ref{reg}) using the mutual information (\ref{h}). For a single interval we obtain
\begin{equation}
H(l)=S_1(l)\,.\label{ocho}
\end{equation}
That is, $H(l)$ is just the single interval entropy (plus an arbitrary additive constant).
Then we can write $C(l)=l\,S_1'(l)$. This indicates that the condition $C^\prime(l)\le 0$ coincides with the entanglement entropy c-theorem for this particular case \cite{cteo,hsh}.  

Using (\ref{ocho}) in (\ref{h}) we obtain the expression for the multicomponent entropy, which coincides with the Hubeny-Rangamani proposal (\ref{confora}). Thus, this is equivalent to extensivity in two dimensions. As we mentioned, this formula holds for the free massless fermion. In contrast, it does not hold for free massive fields. Some numerical results about this fact were reported in \cite{cteo,ch3}. 
 The figure 3 shows two particular examples. 
 These are numerical calculations in the lattice of $I(A,B,C)$ for a Dirac field with different masses. The involved sets are an interval $A$ of size $a$ separated by a distance $d$ from the two adjacent intervals $B$ of length $b$ and $C$ of length $c$. 
 For this configuration $I(A,B,C)$ only involves the mutual information function for two single component sets. 
 Typically, the plotted points correspond to sizes from 200 to 2000 lattice points. The examples shown are for $a=b=c$, $d/a=3$, and $a=d$, $b=c=2/3 a$. The picture shows that the negative sign predominates, but actually both signs of $I(A,B,C)$ are attained by the massive fermion.

 \begin{figure} [tbp]
\centering
\leavevmode
\epsfysize=6.cm
\bigskip
\epsfbox{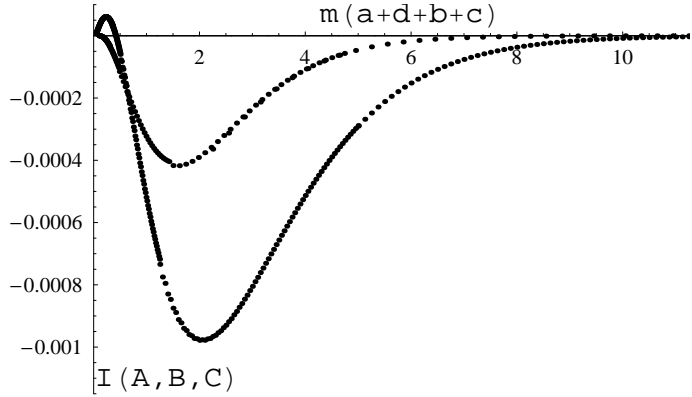}
\caption{ The tripartite information function $I(A,B,C)$ for a two dimensional Dirac field with different masses. The involved sets are an interval $A$ of size $a$ separated by a distance $d$ from the two adjacent intervals $B$ of length $b$ and $C$ of length $c$. The examples shown are for $a=b=c$, $d/a=3$ (the curve with the smaller well), and $a=d$, $b=c=2/3 a$ (the curve at the bottom, with the larger well). The $x$ axis is the product of the mass with the total length $a+d+b+c$.}
\end{figure}
  
We have also done analytical calculations for free massive fermions. These are based on a direct diagonalization of the density matrix for the multicomponent massless case, and perturbations for small mass. The details of the derivation of these analytical results for the deviation from extensivity will be reported elsewhere. Here we just present the result for the mutual information of two intervals $A$ and $B$ of lengths $a$ and $b$ separated by a distance $d\gg a,b$, in the small mass limit, $m d\ll 1$. We have for a Dirac field
\begin{eqnarray}
&&I(A,B)\sim\frac{1}{3}\log\left(\frac{(a+d)(b+d)}{d\, (a+b+d)}\right)+\frac{1}{3}\,a\,b\, m^2 \log^2(m) \\ &&+
\frac{1}{3}\left((2 \log(d)+2\gamma_E -2 \log (2)+ 1) \,a\,b
+\frac{a\,b\,(a+b)}{d}+{\cal O}(d^{-2})\right)\,m^2 \log (m)+{\cal O}(m^2 \log^0(m)) \nonumber
\end{eqnarray}
The first term is the conformal extensive contribution. The second one is the leading correction, which is still extensive. The last term contains the first non-extensive correction. In this case it gives negative tripartite information 
\begin{equation}
I(A,B,C)\sim -\frac{2}{3}\,\,\frac{a\,b\,c}{d}\,m^2 \log (m)\,,
\end{equation}  
where as above $a$ is the size of $A$, and $b$ and $c$ are the ones of the consecutive intervals $B$ and $C$. 

For the free scalar the deviations from extensivity are much larger than for the fermion, and there is no extensivity in the massless limit either. This is because the homogeneous component of the field is a zero mode in this limit. The correlation function diverges logarithmically with the mass, and thus the typical size of the fluctuations on the homogeneous mode grows as $(-\log (m))^{1/2}$. Correspondingly, the entropy grows as the logarithm of this volume in field space \cite{unruh}, and    
 becomes infrared divergent $S(A)\sim 1/2\log(-\log(m))$ \cite{a}. This term in the entanglement entropy is independent of the number of components of $A$ due to its infrared origin. In consequence, $I(A,B)\sim \frac{1}{2}\,\log(-\log(m))$ is also infrared divergent, what impedes the formula (\ref{ex}) to hold in the small mass limit. We have positive tripartite information 
\begin{equation}
I(A,B,C)\sim \frac{1}{2}\,\log(-\log(m))
\end{equation}
in this case. In ref. \cite{hr} the equation (\ref{confora}) was proposed to hold for all two dimensional QFT. As we see, this cannot be the case.

\begin{figure} [tbp]
\centering
\leavevmode
\epsfysize=6cm
\bigskip
\epsfbox{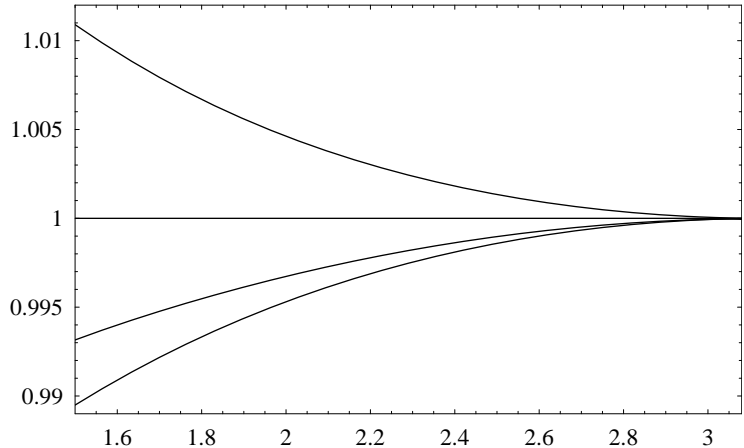}
\caption{Comparison between the coefficients of the logarithmic term in three dimensions for different theories. The picture shows, from top to bottom, the functions $s_S/s_E$, $s_D/s_E$ and $s_H/s_E$. The differences increase toward the origin where these ratios are $s_S/s_E=1.078$, $s_D/s_E=.980$ and $s_H/s_E=.956$ (not shown). The functions $s_H$ and $s_E$ are normalized such that the coefficient of $(x-\pi)^2$ in the Taylor series expansion around $x=\pi$ coincide with the ones of $s_D$ and $s_S$. }
\end{figure}
\subsection{Vertices in three dimensions}
The entanglement entropy $S(V)$ associated to a polygonal set $V$ in three dimensions has a term which runs logarithmically with the cutoff $\epsilon$ 
\begin{equation}
s\log(\epsilon \Lambda)\,, 
\end{equation}
where $\Lambda$ is a parameter with dimension of energy, depending on $V$ and on the details of the theory. The dimensionless universal coefficient $s$, being an extensive and local function on the boundary, is given by 
\begin{equation}
s=\sum_{v_i} s(x_i)\,,
\end{equation} 
where the sum is over the vertices $v_i$ and where $x_i$ is the vertex angle. 
The function $s(x)$ was calculated for a free scalar field $s=s_S$ \cite{b}, and for a Dirac one $s=s_D$ \cite{c}. The calculations are based in the relation between the entropy and a partition function in $D$ dimensions, where the field has particular boundary conditions imposed on the $D-1$ dimensional set $V$ \cite{analitic}.

For the holographic ansatz, a similar term in three dimensions was found by Hirata and Takayanagi \cite{hht}. We call their angular function $s_H(x)$.

On the other hand, from (\ref{h}) and (\ref{reg}) we can calculate the regularized entropy of a plane angular sector and extract the logarithmic contribution to this entropy, obtaining the coefficient $s=s_E$ of the "extensive" function. This is
\begin{equation}
s_E= 2\,C_3 \,(1+(\pi-x)\textrm{cot}(x))\,,
\end{equation} 
with $C_3$ a constant corresponding to the ultraviolet fixed point.

In order to compare the curves of the different angular functions $s(x)$, we normalize $s_H$ and $s_E$ to have the same quadratic coefficient as $s_S$ and $s_D$ in the Taylor series expansion for $x$ around $\pi$. The quadratic coefficients for these later functions coincide. For more details see \cite{c}. The functions $s_D$, $s_S$, $s_E$ and $s_H$, divided by $s_E(x)$, are plotted in figure 4 for a range $x\in [\pi/2,\pi]$. Though they are all remarkably similar to each other, they differ by around $1\%$ in this range. Therefore, we conclude none of these theories is extensive. Moreover, we are able to check that they are linearly independent, using the coefficients of the series expansion around $x=\pi$ reported in \cite{c}. This means there is no non-interacting combination of them which has extensive mutual information. 

\subsection{Dimensional reduction}
The non-extensive character of the mutual information for free massive fields in two and three dimensions is transferred to higher dimensions. In order to see this consider sets $A_{D_1}$, $B_{D_1}$ and $C_{D_1}$ in space-time dimension $D_1$, which are products of $D_2$ dimensional sets $A_{D_2}$, $B_{D_2}$ and $C_{D_2}$ times a large cube of size $L$ in the other $D_1-D_2$ remaining dimensions. Decomposing the $D_1$ dimensional fields in modes with momentum parallel and transversal to the $D_2$ dimension selected, the squared momentum in the transversal direction is interpreted as a squared mass for the $D_2$ dimensional theory. We arrive in this way at a dimensionally reduced formula in the large $L$ limit, which relates the higher dimensional function with the lower dimensional one, summed over different masses \cite{a}
\begin{equation}
I_{D_1}(A_{D_1},B_{D_1},C_{D_1},m)\sim  \frac{q_{D_1}\textrm{V}(S^{D_1-D_2-1})L^{D_1-D_2}}{q_{D_2}(2\pi)^{D_1-D_2}} \int_0^\infty dz\,z^{D_1-D_2-1}\, I_{D_2}(A_{D_2},B_{D_2},C_{D_2},\sqrt{m^2+z^2})\,.\label{this}
\end{equation}
Here $\textrm{V}(S^{(n)})=(n+1)\pi^{(n+1)/2}/\Gamma ((n+3)/2)$ denotes the volume of the $n$ dimensional sphere and $q_n$ is the multiplicity of the spin degree of freedom in $n$ dimensions. 
In particular, when $D_2=D_1-2\equiv D$ we have
\begin{equation}
\frac{\partial}{\partial m}I_{D+2}(A_{D+2},B_{D+2},C_{D+2},m)=-m\frac{q_{D+2}\,L^{2}}{q_{D}\,\,(2\pi)}I_{D}(A_{D},B_{D},C_{D},m)\,.
\end{equation}
This shows explicitly that for a free scalar or Dirac field in any dimension there is a mass such that the theory is non-extensive.   
 Then, there is non extensivity for any nonzero mass, since by dimensional analysis $I(A,B,C,m)\equiv I(\lambda A, \lambda B, \lambda C,m/\lambda)$. For the massless case in more than three dimensions we can use directly (\ref{this}) to relate the tripartite information with the massive lower dimensional ones. The numerical calculations of figure 3 for the fermion fields show that these integrals are not zero in general, and thus the massless fermions are non-extensive in any dimension $D\ge 2$. We will show later the same holds for massless scalars. 
   
Note that (\ref{this}) implies that $I_D(A,B,C,m)$ is not bounded for $D>2$, even if the two dimensional function were bounded.
 
\subsection{Spheres in $D$ dimensions}

Now we consider a conformal theory in $D$ space-time dimensions, and focus attention on the mutual information between a ball bounded by a sphere $A$ of radius $R_A$ and the exterior $B$ of a concentric sphere of radius $R_B$, with $R_B > R_A$. Let us introduce the parameter $x=R_A/R_B$ with $x\in(0,1)$. 
   
The mutual information for the extensive case reads
\begin{eqnarray}
I(A,B)&=&C_D \int d\Omega_A\int d\Omega_B\; R_A^{D-2}R_B^{D-2}\frac{(\vec{\Omega}_A . \vec{\Omega}_B)}{|\bar{\Omega}_AR_A-\bar{\Omega}_BR_B|^{2D-4}}\nonumber \\
&=&C_D\,x^{D-2}\textrm{V}(S^{D-2})\textrm{V}(S^{D-3})\int_0^\pi d\theta\, \cos(\theta)\frac{ \sin(\theta)^{D-3}}{((1+x^2)-2 x \cos(\theta))^{D-2}}\,.
\end{eqnarray}
The result for $D$ odd is  
\begin{equation}
I(A,B)=\frac{\gamma\,\pi\, (-1)^{\frac{D+1}{2}}}{2^{D-3} }\,\,
\,\frac{x^{D-2}\sum_{i=0}^{\frac{D-3}{2}}\, (-1)^{i}\, \binom{D-2}{i}  \,x^{D-2-2 i}}{(1-x^2)^{D-2}}\,,\label{cua}
\end{equation}
with $\gamma=C_D\,\textrm{V}(S^{D-2})\textrm{V}(S^{D-3})$.
For $D$ even we have 
\begin{equation}
I(A,B)=\frac{(-1)^{\frac{D}{2}}\gamma}{2^{D-3} } \,\,\log\left(\frac{1-x}{1+x}\right)+\frac{(-1)^{\frac{D}{2}}\gamma}{2^{D-4}(D-3)!!}\,\,\frac{x^2 \,P_D(x)}{  (1-x^2)^{D-2}}\,,\label{tre}
\end{equation}
where $P_D(x)$ is a polynomial of degree $(2D-6)$. We have not been able to obtain a closed expression of $P_D(x)$ valid for any dimensions. Table 1 shows $P_D(x)$ for $D\le 10$.

\begin{table}[t]
\centering
\begin{tabular}{ c|l } 
  $D$ & $P_D(x)$ \hspace{5cm} \\  \hline
$4$ & $ x^2+1$ \\  
$6$ & $3 x^6 - 11 x^4 -11 x^2 + 3$  \\  
$8$ & $15 x^{10}-85 x^8+ 198 x^6 +198 x^4 -85 x^2 + 15 $ \\ 
$10$ & $105 x^{14}- 805 x^{12} +2681 x^{10} -5053 x^8 -5053 x^6 +2681 x^4 -805 x^2 +105$ \\
\end{tabular} 
\caption{the polynomial $P_D(x)$ for even dimension $D\le 10$.}
\end{table}

Now we can compare these functions with the mutual information corresponding to the Ryu-Takayanagi ansatz. For the present case of two concentric spheres this is given in terms of an integral over the solution of a non linear differential equation \cite{hht}. The important point we want to stress here is that the holographic ansatz displays a "phase transition" (Gross-Ooguri transition \cite{go} in the context of Wilson loop calculations), a non differentiable change in the curve $I(A,B)$ as a function of $x$. This was studied in detail for $D=3$ and $D=4$ in \cite{hht} (see also \cite{al}), but is a general feature for any dimension. This is because the entanglement entropy of an annulus bounded by the two spheres is computed as a minimal area in AdS having the spheres as a boundary. When the spheres are near to each other the topology of this minimal surface is similar to a half torus, directly connecting the surfaces. In contrast, in the case the smaller sphere tends to collapse to a point, the surface of minimal area is necessarily one having two connected components, each one bounded by a different sphere. Then, at some $x\in (0,1)$ there is a change in the topology of the minimal surface, and a non analytical transition. This non analiticity is not present for the functions (\ref{cua}) and (\ref{tre}) corresponding to an extensive mutual information in odd and even dimensions respectively. 
Thus, we conclude that, for any dimension, the entropy function of the Ryu-Takayanagi ansatz cannot correspond to an extensive theory.    

\subsection{Universal coefficients}

The presence of a logarithmic divergent term in the geometric entropy for smooth sets in even dimensions is a general phenomenon. This is a consequence of an anomaly which is present for conical manifolds with curved singularity surface \cite{anoma}.  
In agreement with this fact, the eq. (\ref{tre}) for the extensive case in even dimensions, leads to a logarithmically divergent term in the entropy for a sphere. 
The coefficient of $\log(\epsilon)$, with $\epsilon$ the cutoff, reads 
\begin{equation}
s_E= \frac{(-1)^{\frac{D}{2}}\gamma}{2^{D-2} } \,.
\end{equation} 
For the holographic ansatz the coefficient of the logarithmic term for a sphere in even dimensions is \cite{rt} 
\begin{equation}
s_H=\frac{R^{D-1}}{2G_N^{D+1}}\frac{(-1)^{D/2}\pi^{(D-1)/2}(D-3)!!}{\Gamma\left(\frac{D-1}{2}\right)(D-2)!!}\,.
\end{equation}

It is interesting to note that this coefficient, which in the extensive conformal case is proportional to the function $C$, has been related to the central charges of the theory (coefficients of the energy momentum tensor anomaly) \cite{rt}.  For a sphere in four dimensions the coefficient is proportional to the type $a$ conformal anomaly \cite{so}.

Another interesting universal coefficient can be obtained from the spherical corona in the limit of small separation $x=R_1/R_2\sim 1$.
This is locally equivalent to the case of two sets with large plane parallel faces of area ${\cal A}$ which are near to each other. In this case the associated mutual information has a leading term of the form 
\begin{equation}
I(A,B)\sim k \frac{{\cal A}}{L^{D-2}}\,.
\end{equation}
The coefficient $k$ has been calculated for free fields in \cite{bh,a}.  In the extensive case, we have for odd $D$
\begin{equation}
k_E=(-1)^{D+1} 2^{\frac{7-3D}{2}} \pi \frac{(D-4)!!}{\left(\frac{d-3}{2}\right)!} \frac{\gamma}{\textrm{V}(S^{D-2})}
\,,
\end{equation}
and for $D$ even
\begin{equation}
k_E=\frac{\gamma (D/2-2)!}{2^{D/2-1}(D-3)!!\, \textrm{V}(S^{D-2})}\,.
\end{equation}
On the other hand, for the holographic entropy we have \cite{rt}
\begin{equation}
k_H=\frac{R^{D-1}}{4G_N^{D+1}}\frac{2^{D-1}\pi^{(D-1)/2}}{(D-2)}\left(\frac{\Gamma\left(\frac{D}{2(D-1)}\right)}{\Gamma\left(\frac{1}{2(D-1)}\right)}\right)^{D-1}\,.
\end{equation}

Then, we can compare the ratios of universal coefficients for the holographic and extensive functions in even dimensions
\begin{equation}
\frac{s_H}{k_H} \frac{k_E}{s_E}=\frac{1}{2^{D/2}\pi^{(D-1)/2}}\,\,\left(\frac{\Gamma\left(\frac{1}{2(D-1)}\right)}{\Gamma\left(\frac{D}{2(D-1)}\right)}\right)^{D-1}\,.
\end{equation}
Since this is different from one, it confirms the non-extensive character of the holographic function in this even dimensional case. 

\subsection{Long distance limit}

Let us now look at the long distance limit of the mutual information. From (\ref{jota}) it follows that for $A$ and $B$ contained in the same hyperplane, and in the conformal case, the extensive function is
\begin{equation}
I(A,B)\sim \frac{C(0)}{L^{2(D-1)}}\,\,\,\textrm{Vol}(A) \textrm{Vol}(B)
\,.
\end{equation}  
Note that this expression is invariant under independent rotations of $A$ and $B$, and also independent on the direction of the vector of separation. 

Here we compare this behavior with the case of free massless scalars and fermions. These can be treated using the expression of the local density matrix in terms of correlators \cite{arapes}.  
For a free fermion we have \cite{analitic}
\begin{equation}
S(V)=-\int^\infty_{1/2} d\beta\, \textrm{Tr}\left[\left(\beta-\frac{1}{2}\right) \left(R_V(\beta)-R_V(-\beta)\right)-\frac{2 \beta}{\beta+\frac{1}{2}}\right]\,,
\end{equation}  
where 
\begin{equation}
R_V(\beta)=\left({\cal C}_V+\beta-\frac{1}{2}\right)^{-1}
\end{equation}
is the resolvent of the field correlator ${\cal C}_V$ restricted to $V$, 
\begin{equation}
{\cal C}_V(x,y)=<0|\Psi(x)^\dagger\Psi(y)|0>= \frac{1}{2} \delta(x-y)+i c\, \gamma^i \gamma^0 \frac{(x-y)^i}{|x-y|^{D}}\,,
\end{equation}
where $c$ is a constant depending on the dimension $D$ and $\gamma^\mu$ are the Dirac matrices.

Considering $V=A\cup B$ with a large separation vector $\vec{L}$ between $A$ and $B$, the resolvent can be expanded perturbatively,
\begin{equation}
R_V(\beta)=R_V^0(\beta)-R_V^0(\beta){\cal C}_1 R_V^0(\beta)+R_V^0(\beta){\cal C}_1 R_V^0(\beta){\cal C}_1 R_V^0(\beta)-...\,.
\label{financiar}
\end{equation}
Here $R_V^0(\beta)$ is the unperturbed resolvent, 
\begin{equation}
R_V^0(\beta)=\left(\begin{array}{cc}
R_A(\beta) & 0 \\
0 & R_B(\beta) 
 \end{array}
\right)\,,
\end{equation}
and ${\cal C}_1$ is the field correlator evaluated for the separation vector $\vec{L}$ 
\begin{equation}
{\cal C}_1=i\,c \,\frac{\gamma^i \gamma^0 L^i}{L^{D}} \left(\begin{array}{cc}
0& {\mathbf 1}_{A,B} \\
-{\mathbf 1}_{B,A}& 0 
 \end{array}
\right)\,.
\end{equation}
The kernel ${\mathbf 1}_{A,B}(x,y)=1$ for any $x \in A$ and $y\in B$.

The second term in (\ref{financiar}) does not contribute since it has zero trace. The third term is proportional to the squared of the field correlator and leads to 
\begin{equation}
I(A,B)\sim \frac{c^2}{L^{2(D-1)}}\int^\infty_{1/2} d\beta\,\left(\beta-1/2\right)\left[\, \textrm{tr} \left(  \overline{R_{\hat{A}}}(\beta) \overline{R_B^2}(\beta) +  \overline{R_{\hat{B}}}(\beta)  \overline{R_A^2}(\beta)\right)            -\left(\beta\rightarrow -\beta\right)\right]\,,\label{inn}
\end{equation}
where $\hat{V}$ means the set $V$ after an inversion of coordinates followed by a reflection in the plane perpendicular to $\vec{L}$. The bar over the resolvent and the square of the resolvent means sum over the spatial variables, $\overline{{\cal O}_X}=\int_X dx\, \int_X dy\, {\cal O}(x,y)$.

Thus, the power $I(A,B)\sim L^{-2(D-1)}$ is the same as in the extensive case, but the coefficient may not be invariant under separate or joint  rotations of $A$ and $B$ (it may depend on the direction of $\vec{L}$). Also, the structure of this coefficient would in general be different to the product of the volumes. However, we have not been able to confirm this numerically in $D=3$.

A similar calculation for scalars shows that the mutual information also decays with the field correlator squared. For massless fields this gives $I(A,B)\sim L^{-2(D-2)}$, which is not consistent with extensivity. We can obtain this same result with a different approach. The mutual information is an upper bound on correlation \cite{upper}
\begin{equation}
I(A,B)\ge \frac{1}{2}\frac{\left(\left<O_A O_B\right>-\left<O_A\right>\left<O_B\right>\right)^2}{\left\|O_A\right\|^2\left\|O_B\right\|^2}\label{dfg}\,,
\end{equation}
where $O_A$ and $O_B$ are any bounded operators in the local algebras corresponding to $A$ and $B$, with norms $\left\|O_A\right\|$ and $\left\|O_B\right\|$ respectively. We cannot apply this relation directly to the smeared scalar field operator, $\int dx \,\alpha(x) \phi(x)$, with $\alpha(x)$ an smooth function of compact support, since this is not bounded. However, using the inequality (\ref{dfg}) with the unitary operators $O_A=\exp{(i  \int dx \,\alpha(x) \phi(x))}$ and $O_B=\exp(i \int dy \, \beta(y) \phi(x))$, and $\alpha(x)$, $\beta(x)$ vanishing outside $A$, $B$ respectively, we obtain, 
\begin{equation}
I(A,B)\ge \textrm{cons} \left<\phi(0) \phi(\vec{L})\right>^2\,\sim L^{-2(D-2)}.
\end{equation}
The same argument shows that for any hypothetical extensive theory there can not be correlators for localized  bounded operators falling at a slower rate than $L^{-2(D-1)}$.   

\section{Concluding remarks}
In conclusion, we have disproved extensivity of the mutual information for the Ryu-Takayanagi ansatz, as well as for the entropy functions for free scalar and Dirac fields (excepting the $D=2$ massless Dirac case which is extensive).

The deviations from extensivity, as measured by $I(A,B,C)$, can have any   sign. As far as we have checked, these tend to be relatively small (few percent) for the Dirac field, while they are much greater for the scalar field. The fact that different spins and sets configurations lead to different signs for $I(A,B,C)$ gives some hope that a cancellation may occur for a particular interacting "extensive" theory.

The entanglement entropy offers a non standard view of some aspects of QFT, and one can even hope that it might provide an alternative setting where QFT may be expressed in a geometrical way.
 An early attempt pursued this idea from an axiomatic point of view \cite{hh}, looking for the entropy functions $S(V)$ which are subjected to the very general conditions of relativistic invariance, causality, positivity and strong subadditivity. The result can be reworded in the following terms. One either has divergent entanglement entropy, or the mutual information is zero. This last option is clearly inappropriate, since it implies the vanishing of all the correlators.    
 
Facing the fact that the entropy must diverge in a relativistic theory, one is lead to consider its universal part, which is encoded in the mutual information $I(A,B)$. A list of requirements on this function is as follows:

\noindent a) It is a symmetric ($I(A,B)=I(B,A)$)
function on the pairs of spatially separated causally complete sets in Minkowki space-time 
 
\noindent b) Poincar\'e invariance
 
\noindent c) Positivity  

\noindent d) It is monotonically increasing under inclusion

\noindent e) $I(A,B+\vec{x})$ goes to zero for large translations  $|\vec{x}|\rightarrow \infty$

\noindent f) $I(A,B,C)\equiv I(A,B)+I(A,C)-I(A,B\cup C)$ is invariant under permutation of $A$, $B$ and $C$.

\noindent g) For two identical $D-1$ dimensional cubes $A$ and $B$ (or equivalently their causal completion) with parallel faces, separated by a distance $L$, I(A,B) diverges as $L^{-(D-2)}$ (or $\log(L)$ for $D=2$) in the short $L$ limit.

We have added the condition (g) in order to allow only for solutions representing QFT with a finite number of fields. If this is not imposed one can construct solutions by dimensional reduction from higher dimension, as was done in Section 4. These types of functions have a more singular short distance behavior than the one given by (g) \footnote{This is also the case of the relation between theories of different dimensionality in the Rehren AdS-CFT construction \cite{reh}. In this case the theories related have the same type of short distance divergence of the mutual information, meaning that at least one of them has the wrong behavior for its corresponding dimension.}.

Of course, any QFT gives a solution of the above (a-g) items. An interesting question is the converse. The expression (\ref{h}) with a positive, decreasing $C(l)$, gives the first known explicit solution to this set of constraints. Another meaningful question is how to complete the holographic ansatz in order to fulfill these requirements.

\section*{Acknowledgments}
H.C. and M.H. thank CONICET, ANPCyT and UNCuyo for financial support.

\end{document}